\documentclass[proceedings]{stacs}
\stacsheading{2009}{63--74}{Freiburg}
\firstpageno{63}

\usepackage[linesnumbered,noend]
           {algorithm2e} 
\usepackage{amssymb}     
\usepackage[usenames]
           {color}       
\usepackage{comment}     
\usepackage{enumerate}   
\usepackage{epsfig}      
\usepackage{graphicx}    
\usepackage{latexsym}    
\usepackage{subfigure}   
\usepackage{url}         

\usepackage[
  a4paper,
  linktocpage=true,      
  colorlinks=true,       
  pdftitle={Shortest Paths Avoiding Forbidden Subpaths},
  pdfauthor={Mustaq Ahmed, Anna Lubiw},
  pdfsubject={Computer Science},
  pdfkeywords={Algorithms and data structures,
               Graph algorithms,
               Optical networks,
               Shortest paths,
               Dijkstra,
               Vertex replication}
  ]{hyperref}

\newtheorem*{lemma*}{Lemma}

\begin{document}

\title{Shortest Paths Avoiding Forbidden Subpaths}
\author[UW]{M. Ahmed}{Mustaq Ahmed}
\author[UW]{A. Lubiw}{Anna Lubiw}
\address[UW]{
  David R. Cheriton School of Computer Science,
  University of Waterloo,
  Canada
}
\email{m6ahmed@uwaterloo.ca}
\email{alubiw@uwaterloo.ca}
\thanks{Research partially supported by Nortel Networks and NSERC}


\begin{abstract}
%

%

In this paper we study a variant of the shortest path problem in
graphs: given a weighted graph $G$ and vertices $s$ and $t$, and given
a set $X$ of forbidden paths in $G$, find a shortest $s$-$t$ path $P$
such that no path in $X$ is a subpath of $P$. Path $P$ is allowed to
repeat vertices and edges. We call each path in $X$ an
\emph{exception}, and our desired path a \emph{shortest
exception avoiding path}. We formulate a new version of the problem
where the algorithm has no a priori knowledge of $X$, and finds out
about an exception $x \in X$ only when a path containing $x$
fails. This situation arises in computing shortest paths in optical
networks. We give an algorithm that finds a shortest exception
avoiding path in time polynomial in $|G|$ and $|X|$. The main idea is
to run Dijkstra's algorithm incrementally after replicating vertices
when an exception is discovered.

\end{abstract}

\maketitle

%

%

\section{Introduction}
\label{L2:Intro}

One of the most fundamental combinatorial optimization problems is
that of finding shortest paths in graphs. In this paper we study a
variant of the shortest path problem: given a weighted graph $G(V,
E)$, and vertices $s$ and $t$, and given a set $X$ of \emph{forbidden
paths\/} in $G$, find a shortest $s$-$t$ path $P$ such that no path in
$X$ is a subpath of $P$. We call paths in $X$ \emph{exceptions}, and
we call the desired path a \emph{shortest exception avoiding path}. We
allow an exception avoiding path to be non-simple, i.e.,~to repeat
vertices and edges. In fact the problem becomes hard if the solution
is restricted to simple paths~\cite{Szeider.03}. This problem has been
called the \emph{Shortest Path Problem with Forbidden Paths\/} by
Villeneuve and Desaulniers~\cite{Villeneuve.05}. Unlike them, we
assume no a priori knowledge of $X$. More precisely,
we can identify a forbidden path only after failing in our attempt to
follow that path. This variant of the problem has not been studied
before. It models the computation of shortest paths in optical
networks, described in more detail in the ``Motivation'' section
below. Note that when we fail to follow a path because of a newly
discovered exception, we are still interested in a shortest path from
$s$ to $t$ as opposed to a detour from the failure point. This is what
is required in optical networks, because intermediate nodes do not
store packets, and hence $s$ must resend any lost packet.

This paper presents two algorithms to compute shortest
exception avoiding paths in the model where exceptions are not known a
priori. The algorithms take respectively $O(k n \log n + k m) $ and
$O((n + L) \log (n + L) + m + d L)$ time to find shortest exception
avoiding paths from $s$ to all other vertices, where
$n = |V|$, $m = |E|$,
$d$ is the largest degree of a vertex,
$k$ is the number of exceptions in $X$, and
$L$ is the total size of all exceptions.

Our algorithm uses a vertex replication technique similar to the one
used to handle non-simple paths in other shortest path
problems~\cite{Martins.84,Villeneuve.05}. The idea is to handle a
forbidden path by replicating its vertices and judiciously deleting
edges so that one copy of the forbidden path is missing its last edge
and the other copy is missing its first edge. The result is to exclude
the forbidden path but allow all of its subpaths. The main challenge
is that vertex replication can result in an exponential number of
copies of any forbidden path that overlaps the current one. Villeneuve
and Desaulniers~\cite{Villeneuve.05} address this challenge by
identifying and compressing the overlaps of forbidden paths, an
approach that is impossible for us since we do not have access to
$X$. Our new idea is to couple vertex replication with the ``growth''
of a shortest path tree. By preserving certain structure in the
shortest path tree we prove that the extra copies of forbidden paths
that are produced during vertex replication are immaterial. Our
algorithm is easy to implement, yet the proof of correctness and the
run-time analysis are non-trivial.

\subsection{Motivation}
\label{L3:Motivation}

Our research on shortest exception avoiding path was motivated by a
problem in optical network routing from Nortel Networks. In an optical
network when a ray of light of a particular wavelength tries to follow
a path $P$ consisting of a sequence of optical fibers, it may fail to
reach the endpoint of $P$ because of various transmission impairments
such as attenuation, crosstalk, dispersion and
non-linearities~\cite{Gouveia.03,Lee.05}.  This failure may happen
even though the ray is able to follow \emph{any} subpath $P'$ of
$P$. This non-transitive behavior occurs because those impairments
depend on numerous physical parameters of the traversed path
(e.g.,~length of the path, type of fiber, wavelength and type of laser
used, location and gain of amplifiers, number of switching points,
loss per switching point, etc.), and the effect of those parameters
may be drastically different in $P$ than in
$P'$~\cite{Ashwood-smith.07}. Forbidden subpaths provide a
straight-forward model of this situation.

We now turn to the issue of identifying forbidden paths. Because of
the large number of physical parameters involved, and also because
many of the parameters vary over the lifetime of the
component~\cite{Ashwood-smith.05}, it is not easy to model the
feasibility of a path. Researchers at Nortel suggested a model whereby
an algorithm identifies a potential path, and then this path is tried
out on the actual network. In case of failure, further tests can be
done to pinpoint a minimal forbidden subpath. Because such tests are
expensive, a routing algorithm should try out as few paths as
possible. In particular it is practically impossible to identify all
forbidden paths ahead of time---we have an exponential number of
possible paths to examine in the network. This justifies our
assumption of having no a priori knowledge of the forbidden paths, and
of identifying forbidden paths only by testing feasibility of a path.

The shortest exception avoiding path problem may also have application
in vehicle routing. Forbidden subpaths involving pairs of edges occur
frequently (``No left turn'') and can occur dynamically due to rush
hour constraints, lane closures, construction, etc. Longer forbidden
subpaths are less common, but can arise, for example if heavy traffic
makes it impossible to turn left soon after entering a multi-lane
roadway from the right. If we are routing a single vehicle it is more
natural to find a detour from the point of failure when a forbidden
path is discovered. This is different from our model of rerouting from
$s$ upon discovery of a forbidden path. However, in the situation when
vehicles will be dispatched repeatedly, our model does apply.

\subsection{Preliminaries}
\label{L3:Prelim}

We are given an directed graph $G(V,E)$ with $n = |V|$ vertices and $m
= |E|$ edges where each edge $e \in E$ has a positive weight denoting
its \emph{length}. We are also given a source vertex $s \in V$, a
destination vertex $t \in V$, and a set $X$ of paths in $G$. The graph
$G$ together with $X$ models a communication network in which a packet
cannot follow any path in $X$ because of the physical constraints
mentioned in Sec.~\ref{L3:Motivation}. We assume that the algorithm
can access the set $X$ of forbidden paths only by performing queries
to an oracle. Each query is a path $P$, and the oracle's response is
either the confirmation that $P$ is exception avoiding, or else an
exception $x \in X$ that is a subpath of $P$ and whose last vertex is
earliest in $P$. Ties can be broken arbitrarily. In our discussion we
say ``we try a path'' instead of saying ``we query the oracle''
because the former is more intuitive. In Sec.~\ref{L2:MPDijk2} we
modify our algorithm for the case of an oracle that returns
\emph{any\/} exception on a path (not just the one that ends
earliest). This requires more calls to the oracle but gives a faster
run-time.

We want to find a shortest path from $s$ to $t$ that does not contain
any path in $X$ as a subpath---we make the goal more precise as
follows. A \emph{path} is a sequence of vertices each joined by an
edge to the next vertex in the sequence. Note that we allow a path to
visit vertices and edges more than once. If a path does not visit any
vertex more than once, we explicitly call it a \emph{simple path}. A
simple directed path from vertex $v$ to vertex $w$ in $G$ is called a
\emph{forbidden path} or an \emph{exception} if a packet cannot follow
the path from $v$ to $w$ because of the physical constraints. Given a
set $A$ of forbidden paths, a path $(v_1, v_2, v_3, \ldots, v_l)$ is
said to \emph{avoid} $A$ if $(v_i, v_{i+1}, \ldots v_j) \not\in A$ for
all $i,j$ such that $1 \le i < j \le l$. A path $P$ from $s$ to $t$ is
called a \emph{shortest $A$-avoiding path} if the length of $P$ is the
shortest among all $A$-avoiding paths from $s$ to $t$. We will use the
term ``exception avoiding'' instead of ``$X$-avoiding'' when $A$ is
equal to $X$, the set of all forbidden paths in $G$.

\subsection{Related work}
\label{L3:Related}

A shortest $s$-$t$ path in a graph can be computed in $O(n\log n + m)$
time and linear space using Dijkstra's algorithm with Fibonacci heaps
if all edge weights are non-negative, and in $O(m n)$ time and linear
space using the Bellman-Ford algorithm
otherwise~\cite{Cormen.01}. When the edge weights are non-negative
integers, the problem can be solved in deterministic $O(m \log\log n
\log\log\log n)$ time and linear space if the graph is
directed~\cite{Han.01}, and in optimal $O(m)$ time if the graph is
undirected~\cite{Thorup.99}. In many of these cases, there are
randomized algorithms with better expected times as well as
approximation schemes. See Zwick~\cite{Zwick.01} for a survey of
shortest path algorithms, and Cabello~\cite{Cabello.06}, Goldberg and
Harrelson~\cite{Goldberg.05} and Holzer et al.~\cite{Holzer.05} for
some of the more recent work.

Two recent papers on shortest paths in graphs address the issue of
avoiding a set of forbidden paths, assuming that all the forbidden
paths are known a priori. The first paper gives a hardness
result. Szeider~\cite{Szeider.03} shows, using a reduction from 3-SAT,
that the problem of finding a shortest \emph{simple\/} exception
avoiding path is NP-complete even when each forbidden path has two
edges. If the forbidden paths are \emph{not\/} known a priori, the
hardness result still applies to the case of simple paths because the
lack of prior knowledge of the forbidden paths only makes the problem
harder.

The second paper, by Villeneuve and Desaulniers~\cite{Villeneuve.05},
gives an algorithm for a shortest (possibly non-simple) exception
avoiding path for the case when all the forbidden paths are known a
priori. They preprocess the graph in $O((n+L) \log(n+L) + m + d L)$
time and $O(n + m + d L)$ space so that a shortest path from $s$ to a
query vertex can be found in $O(n + L)$ time. They first build a
deterministic finite automaton (DFA) from the set of forbidden paths
using the idea of Aho and Corasick~\cite{Aho.75}, which can detect in
linear time whether a given path contains any of the forbidden
paths. They then ``insert'' the DFA into $G$ by replicating certain
vertices of $G$ in the manner introduced by Martins~\cite{Martins.84},
and then build a shortest path tree in this modified graph.
Their algorithm cannot handle the case where the set of all forbidden
paths is not explicitly given. Our algorithm is strictly more general,
and we show in Sec.~\ref{L2:MPDijk2} that it solves their problem in
roughly the same time but in less ($O(n + m + L)$) space.

We now mention two problems that seem related to ours, but do not in
fact provide solutions to ours. The first one is maintaining shortest
paths in a dynamic graph, i.e.,~where nodes or edges may
fail~\cite{Demetrescu.04,Demetrescu.07,Hershberger.07}, or edge
weights may change
(e.g.,~\cite{Demetrescu.04,Demetrescu.01}). Forbidden paths cannot be
modeled by deleting edges or by modifying edge costs because
\emph{all\/} edges in a particular forbidden path may be
essential---see Fig.~\ref{fig:NoSPTree} for an example. The second
seemingly related problem is finding the $k$ shortest paths in a
graph. This was the subject of Martins~\cite{Martins.84} who
introduced the vertex replication technique that we use in our
algorithm. There is considerable work on this problem, see Eppstein
~\cite{Eppstein.99} for a brief survey. But the $k$ shortest path
problem is again different from our situation because a forbidden
subpath may be a bottleneck that is present in all of the $k$ shortest
paths even for $k \in \Omega(2^{n/2})$, see Villeneuve and
Desaulniers~\cite{Villeneuve.05}.

In the context of optical networks researchers have studied many
theoretical problems. See Ramaswami and Sivarajan~\cite{Ramaswami.02}
for details on optical networks, and Lee and Shayman~\cite{Lee.05} and
McGregor and Shepherd~\cite{McGregor.07} for a brief survey of the
theoretical problems that have been investigated. In the previous
work, the effect of physical constraints on paths in optical networks
is either not considered at all (e.g.,~Khuller et
al.~\cite{Khuller.05}), or simply modeled by a known constant upper
bound on the length of such a path (e.g.,~Gouveia et
al~\cite{Gouveia.03}, Lee and Shayman~\cite{Lee.05} and McGregor and
Shepherd~\cite{McGregor.07}). To the best of our knowledge, none of
the previous work on shortest paths in optical networks considers the
fact that it is practically infeasible to know a priori all the
forbidden paths in the network, i.e.,~all the constraints in $X$. Our
paper handles the issue of physical constraints from a different and
much more practical perspective.

%

%

\section{Algorithm for a shortest \texorpdfstring{$s$-$t$}{s-t} path}
\label{L2:MPDijk}

\begin{figure}[tb]
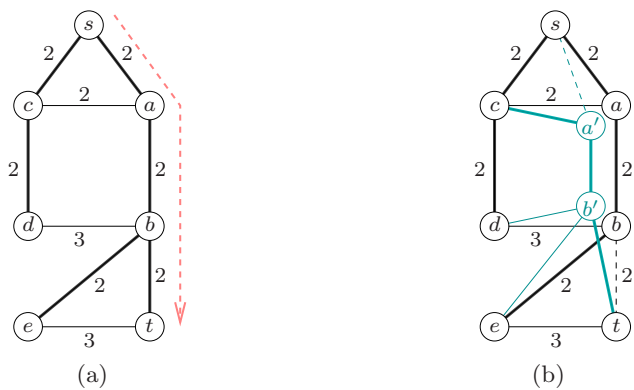

  \hspace*{\fill}
  \subfigure[]{
    \input{fig/NoSPTree-1.pstex_t}
    \label{fig:NoSPTree-1}
  }
  \hspace*{\fill}
  \subfigure[]{
    \input{fig/NoSPTree-2.pstex_t}
    \label{fig:NoSPTree-2}
  }
  \hspace*{\fill}

  \vspace{-1.2ex}
  \caption{(a) Shortest paths and (b) shortest $x$-avoiding paths
    in a graph, where $x = (s, a, b, t)$.}
  \label{fig:NoSPTree}
\end{figure}

In our algorithm we begin with a shortest path tree rooted at $s$,
ignoring the exceptions. We then ``try out'' the path from $s$ to $t$
in the tree. If the path is free of exceptions, we are
done. Otherwise, to take the newly discovered exception into account,
we modify the graph using path replication as described in the
Introduction, and we modify the shortest path tree to match. In
general, we maintain a modified graph and a shortest path tree in the
graph that gives a shortest path in the original graph from $s$ to
every other vertex avoiding all the currently-known exceptions. We
will first illustrate the idea with an example. Consider the graph $G$
in Fig.~\ref{fig:NoSPTree-1}, where the integers denote edge weights,
and the dashed arrow marks the forbidden path $x = (s, a, b, t)$. Note
that for simplicity we have used undirected edges in the figure to
denote bidirectional edges. It is not hard to see that $P= (s, c, a,
b, t)$ is the shortest $x$-avoiding path from $s$ to $t$. To find $P$,
we first construct a shortest path tree rooted at $s$ (marked using
the heavy edges in Fig.~\ref{fig:NoSPTree-1}), and then try the path
$(s, a, b, t)$ in the tree. The path fails because it contains $x$, so
we use a \emph{vertex replication technique} similar to the one by
Martins~\cite{Martins.84} to make duplicates of vertices $a$ and $b$
and delete edges $(s, a')$ and $(b, t)$, as shown in
Fig.~\ref{fig:NoSPTree-2}. We then construct a shortest path tree
rooted at $s$ (marked using the heavy edges in
Fig.~\ref{fig:NoSPTree-2}) in the modified graph, and try the path
$(s, c, a', b', t)$ which ``represents'' the path $P$ in $G$. We are
done if $x$ is the only forbidden path in $G$. Note that this approach
can double the number of \emph{undiscovered} forbidden paths. Suppose
$y = (c, a, b)$ is another forbidden path in $G$. We have two copies
of $y$ in the modified graph: $(c,a,b)$ and $(c,a',b')$, and we have
to avoid both of them. Our solution to this doubling problem is to
``grow'' the shortest path tree in such a way that at most one of
these two copies is encountered in future. Our algorithm is as
follows:

\begin{algorithm}[H]
  construct the shortest path tree $T_0$ rooted at $s$ in $G_0 = G$\;
  \nllabel{alg.Start}
  
  let $i = 1$\;
  
  send a packet from $s$ to $t$ through the path in $T_0$\;

  \While{the packet fails to reach $t$\nllabel{alg.WhileBegin}}
  {
    let $x_i$ be the exception that caused the failure\;
    \nllabel{alg.IndentifyX}

    construct $G_i$ from $G_{i-1}$ by replicating the intermediate
    vertices of $x_i$ and then deleting selected edges\;
    \nllabel{alg.ModifyG}

    construct the shortest path tree $T_i$ rooted at $s$ in $G_i$
    using $T_{i-1}$\;
    \nllabel{alg.MakeT}

    send a packet from $s$ to $t$ through the path in $T_i$\;
    \nllabel{alg.Send}

    let $i = i+1$\;
	
  }
\end{algorithm}

In the above algorithm, the only lines that need further discussion
are Lines~\ref{alg.ModifyG} and~\ref{alg.MakeT}; details are in
Sections~\ref{L3:MPDijk.ModifyG} and~\ref{L3:MPDijk.MakeT}
respectively. In the rest of the paper, whenever we focus on a
particular iteration $i > 0$, we use the following notation:
(i) the path from $s$ to $t$ in $T_{i-1}$, i.e.,~the path along which
we try to send the packet to $t$ in Line~\ref{alg.WhileBegin} in the
iteration, is $(s, v_1, v_2, \ldots, v_p, t)$, and
(ii) the exception that prevented the packet from reaching $t$ in the
iteration is $x_i = (v_{r-l}, v_{r-l+1}, \ldots, v_r, v_{r+1})$, which
consists of $l + 1$ edges.

\subsection{Modifying the graph}
\label{L3:MPDijk.ModifyG}

The modification of $G_{i-1}$ into $G_i$ (Line~\ref{alg.ModifyG}) in
the $i$th iteration eliminates exception $x_i$ while preserving all
the $x_i$-avoiding paths in $G_{i-1}$. We do the modification in two
steps.

In the first step, we create a graph $G'_{i-1}$ by replicating the
intermediate vertices of $x_i$ (i.e.,~the vertices $v_{r-l+1},
v_{r-l+2}, \ldots, v_r$). We also add appropriate edges to the replica
$v'$ of a vertex $v$. Specifically, when we add $v'$ to $G_{i-1}$, we
also add the edges of appropriate weights between $v'$ and the
neighbors of $v$. It is easy to see that if a path in $G_{i-1}$ uses
$l' \le l$ intermediate vertices of $x_i$, then there are exactly
$2^{l'}$ copies of the path in $G'_{i-1}$. We say that a path in
$G'_{i-1}$ is $x_i$-avoiding if it contains none of the $2^l$ copies
of $x_i$.

\begin{figure}[tb]
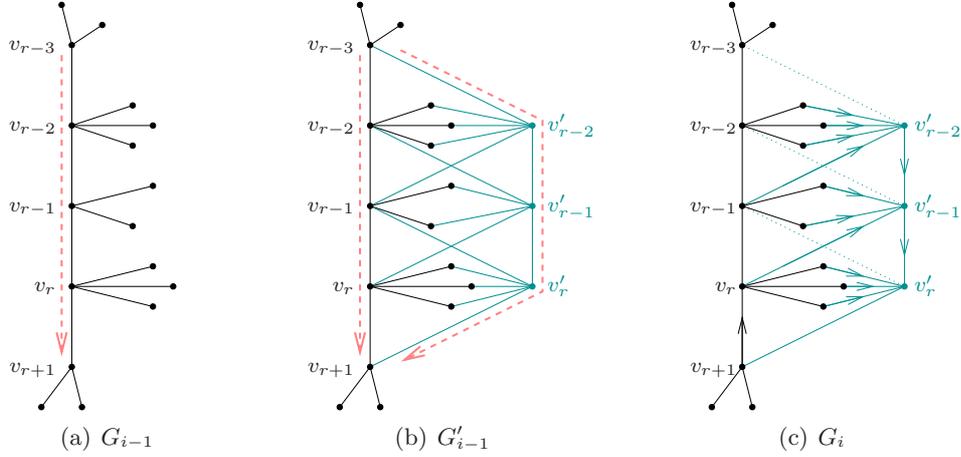

  \hspace*{\fill}
  \subfigure[$G_{i-1}$]{
    \input{fig/ModifyG-1.pstex_t}
    \label{fig:ModifyG-1}
  }
  \hspace*{\fill}
  \subfigure[$G'_{i-1}$]{
    \input{fig/ModifyG-2.pstex_t}
    \label{fig:ModifyG-2}
  }
  \hspace*{\fill}
  \subfigure[$G_i$]{
    \input{fig/ModifyG-3.pstex_t}
    \label{fig:ModifyG-3}
  }
  \hspace*{\fill}

  \vspace{-1.2ex}
  \caption{Modifying $G_{i-1}$ to $G_i$: (a) The part of $G_{i-1}$ at
    an exception $(v_{r-3}, v_{r-2}, v_{r-1}, v_r, v_{r+1})$, with $l
    = 3$. (b) Replicating vertices to create $G'_{i-1}$. The dashed
    paths show two of the $8$ copies of the exception. (c) Deleting
    edges to create $G_i$. The dotted lines denote deleted edges. }
  \label{fig:ModifyG}
\end{figure}

In the second step, we build a spanning subgraph $G_i$ of $G'_{i-1}$
by deleting a few edges from $G'_{i-1}$ in such a way that all copies
of $x_i$ in $G'_{i-1}$ are eliminated, but all $x_i$-avoiding paths in
$G'_{i-1}$ remain unchanged. To build $G_i$ from $G'_{i-1}$, we delete
the edges $(v_{j-1}, v'_j)$ and $(v'_j, v_{j-1})$ for all $j \in
[r-l+1, r]$. We also delete the edge $(v_r, v_{r+1})$, all the
outgoing edges from $v'_r$ except $(v'_r,v_{r+1})$, and all the
outgoing edges from $v'_j$ except $(v'_j, v'_{j+1})$ for all $j \in
[r-l+1, r-1]$. Figure~\ref{fig:ModifyG} shows how the ``neighborhood''
of an exception changes from $G_{i-1}$ to $G_i$. As
before, the undirected edges in the figure are bidirectional.

\begin{observation}\label{obs:GiHasNoX}
  Graph $G_i$ has no copy of $x_i$.
\end{observation}


In Sec.~\ref{L3:MPDijkC.ModifyGProof} we will prove that $G_i$ still
contains all the $x_i$-avoiding paths of $G_{i-1}$.

The vertices in $G_i$ [$G'_{i-1}$] that exist also in $G_{i-1}$
(i.e.,~the ones that are not replica vertices) are called the
\emph{old vertices of $G_i$ [respectively $G'_{i-1}$]}. Note that the
vertices of $G_0$ exist in $G_i$ for all $i \ge 0$. These vertices are
called the \emph{original vertices of $G_i$}.

\subsection{Constructing the tree}
\label{L3:MPDijk.MakeT}

In Line~\ref{alg.MakeT} of our algorithm we construct a tree $T_i$
that contains a shortest $x_i$-avoiding path from $s$ to every other
vertex in $G_{i-1}$. Tree $T_i$ is rooted at $s$, and its edges are
directed away from $s$. Not every shortest path tree rooted at $s$ in
$G_i$ will work. In order to guarantee termination of the algorithm,
$T_i$ must be similar to $T_ {i-1}$, specifically, every
$x_i$-avoiding path from $s$ in $T_{i-1}$ must be present in
$T_i$. The necessity of this restriction is explained in
Sec.~\ref{L3:MPDijkC.Analysis}.

We construct the required $T_i$ by preserving as much of $T_{i-1}$ as
possible. We apply Dijkstra's algorithm starting from the part of
$T_{i-1}$ that can be preserved. Let $V'$ be the set of vertices that
are either replica vertices in $G_i$, or descendants of $v_{r+1}$ in
$T_{i-1}$. We first set the weight of each $v \in V'$ to infinity, and
temporarily set $T_i = T_{i-1} - V'$. Then, for each $v \in V'$, we
set the weight of $v$ to the minimum, over all edges $(u, v)$, of the
sum of the weight of $u$ and the length of $(u,v)$. Finally, we
initialize the queue used in Dijkstra with all the vertices in $V'$
and run the main loop of Dijkstra's algorithm. Each iteration of the
loop adds one vertex in $V'$ to the temporary $T_i$. When the queue
becomes empty, we get the final tree $T_i$.

\section{Correctness and analysis}
\label{L2:MPDijkC}

\subsection{Justifying the graph modification}
\label{L3:MPDijkC.ModifyGProof}

In this section we prove the following lemma, which uses the notion of
a \emph{corresponding path}. Consider any path $P_i$ in $G_i$. By
substituting every vertex in $P_i$ that is not present in $G_{i-1}$
with the corresponding old vertex in $G_{i-1}$, we get the
corresponding path $P_{i-1}$ in $G_{i-1}$. This is possible because
any ``new'' edge in $G_i$ is a replica of an edge in $G_{i-1}$. We
define the corresponding path $P_j$ in $G_j$ for all $j < i$ by
repeating this argument.

\def\lemShortestInGk{
  If $P_i$ is a shortest path from $s$ to an original vertex $v$ in
  $G_i$, $P_0$ is a shortest $\{x_1, x_2, \ldots, x_i\}$-avoiding
  path from $s$ to $v$ in $G_0$.
}

\begin{lemma*}
  \lemShortestInGk
\end{lemma*}

To prove the above lemma (repeated as Lemma~\ref{lem:ShortestInGk}
below), we will first prove that $x_i$-avoiding paths in $G_{i-1}$ are
preserved in $G_i$ (Lemma~\ref{lem:OldPathInGi}), using the following
characteristic of an $x_i$-avoiding path in the intermediate graph
$G'_{i-1}$:

\begin{lemma}\label{lem:OldPathInGprime}
  For any $x_i$-avoiding path $P$ from $s$ to $v$ that uses only the
  old vertices in $G'_{i-1}$, there exists a copy of $P$ in $G_i$ that
  starts and ends at the old vertices $s$ and $v$ respectively, and
  possibly passes through the corresponding replicas of its
  intermediate vertices.
\end{lemma}


\begin{proof}
  Graph $G_i$ contains all the edges between pairs of old vertices in
  $G'_{i-1}$ except for the directed edge $(v_r, v_{r+1})$. Thus $P$
  can remain unchanged if it does not use this directed
  edge. Otherwise we will re-route any portion of $P$ that uses the
  directed edge $(v_r, v_{r+1})$ to use the replica edge $(v'_r,
  v_{r+1})$ instead. Let $P = (s=w_1, w_2, \ldots, w_q=v)$, and $(w_j,
  w_{j+1})$ be an occurrence of $(v_r, v_{r+1})$ in $P$. Tracing $P$
  backwards from $w_j$, let $h \le j$ be the minimum index such that
  $(w_h, w_{h+1}, \ldots, w_{j+1})$ is a subpath of $x_i$. Because $P$
  is $x_i$-avoiding, $w_h$ must be an intermediate vertex of
  $x_i$. This implies that $h > 1$, since $s = w_1$ is not an
  intermediate vertex of $x_i$ because of the following reasons: (i)
  $x_i$ is a path in the shortest path tree rooted at $s$ in $G_i$,
  and (ii) there is no replica of $s$ in $G_i$. Therefore $w_{h-1}$
  exists. We will reroute the portion of $P$ between $w_{h-1}$ and
  $w_{j+1}$ by using the corresponding replica vertices in place of
  the subpath $(w_h, \ldots, w_j)$ of $x_i$. Note that the required
  edges exist in $G_i$ (since $P$ does not contain the whole exception
  $x_i$), and that the portions of $P$ that we re-route are disjoint
  along $P$. Moreover, $P$ starts and ends at the old vertices $s$ and
  $v$ respectively.
\end{proof}

\begin{lemma}\label{lem:OldPathInGi}
  Any $x_i$-avoiding path from $s$ to $v$ in $G_{i-1}$ has a copy in
  $G_i$ that starts and ends at the old vertices $s$ and $v$
  respectively, and possibly goes through the corresponding replicas
  of its intermediate vertices.
\end{lemma}

\begin{proof}
  Let $P$ be the $x_i$-avoiding path in $G_{i-1}$. As we do not delete
  any edge to construct $G'_{i-1}$ from $G_{i-1}$, $P$ remains
  unchanged in $G'_{i-1}$. Moreover, $P$ uses no replica vertex in
  $G'_{i-1}$. So, Lemma~\ref{lem:OldPathInGprime} implies that $P$
  exists in $G_i$ with the same old vertices at the endpoints,
  possibly going through the corresponding replicas of the
  intermediate vertices.
\end{proof}


\begin{lemma}\label{lem:ShortestInGk}
  \lemShortestInGk
\end{lemma}

\begin{proof}
  For any $j \in [0,i]$, let $X_j = \{ x_{j+1}, x_{j+2}, \ldots, x_i
  \}$. We show that for any $j$, if $P_j$ is a shortest $X_j$-avoiding
  path in $G_j$, then $P_{j-1}$ is a shortest $X_{j-1}$-avoiding path
  in $G_{j-1}$. The lemma then follows by induction on $j$, with basis
  $j = i$, because $X_i = \emptyset$ and thus $P_i$ is a shortest
  $X_i$-avoiding path in $G_i$.

  If $P_j$ is a shortest $X_j$-avoiding path in $G_j$, $P_j$ is
  $X_{j-1}$-avoiding because $P_j$ is $x_j$-avoiding by
  Observation~\ref{obs:GiHasNoX}, and $X_j \cup \{x_j\} =
  X_{j-1}$. So, the corresponding path $P_{j-1}$ is also
  $X_{j-1}$-avoiding. If we assume by contradiction that $P_{j-1}$ is
  not a shortest $X_{j-1}$-avoiding path in $G_{j-1}$, then there
  exists another path $P'_{j-1}$ from $s$ to $v$ in $G_{j-1}$ which is
  $X_{j-1}$-avoiding and is shorter than $P_{j-1}$. Since $x_j \in
  X_{j-1}$, $P'_{j-1}$ is $x_j$-avoiding, and hence by
  Lemma~\ref{lem:OldPathInGi}, there is a copy $P'_j$ of path
  $P'_{j-1}$ in $G_j$ which has the same original vertices at the
  endpoints. As $P'_{j-1}$ is $X_j$-avoiding, $P'_j$ is also
  $X_j$-avoiding. This is impossible because $P'_j$ is shorter than
  $P_j$. Therefore, $P_{j-1}$ is a shortest $X_{j-1}$-avoiding path in
  $G_{j-1}$.
\end{proof}

\subsection{Justifying the tree construction}
\label{L3:MPDijkC.MakeTProof}

To show that the ``incremental'' approach used in
Sec.~\ref{L3:MPDijk.MakeT} to construct $T_i$ is correct, we first
show that the part of $T_{i-1}$ that we keep unchanged in $T_i$ is
composed of shortest paths in $G_i$:

\begin{lemma}\label{lem:OldPathInTi}
  For every vertex $v$ that is not a descendant of $v_{r+1}$ in
  $T_{i-1}$, the path $P$ from $s$ to $v$ in $T_{i-1}$ is a shortest
  path in $G_i$.
\end{lemma}

\begin{proof}
  First we show that $P$ exists in $G_i$. Every vertex in $T_{i-1}$
  exists in $G_i$ as an old vertex. So, $P$ exists in $G_i$ through
  the old vertices if no edge of $P$ gets deleted in $G_i$. The only
  edge between a pair of old vertices in $G_{i-1}$ that gets deleted
  in $G_i$ is $(v_r, v_{r+1})$. Since $v$ is not a descendant of
  $v_{r+1}$ in $T_{i-1}$, $P$ does not use the edge $(v_r,
  v_{r+1})$. Therefore, no edge of $P$ gets deleted in $G_i$. So, $P$
  exists in $G_i$ through the old vertices.

  Neither the modification from $G_{i-1}$ to $G'_{i-1}$ nor the one
  from $G'_{i-1}$ to $G_i$ creates any ``shortcut'' between any pair
  of vertices. So, there is no way that the distance between a pair of
  old vertices decreases after these modifications. Since these
  modifications do not change $P$, which is a shortest path in
  $G_{i-1}$, $P$ is a shortest path in $G_i$.
\end{proof}

\begin{lemma}\label{lem:TiIsSPTree}
  The tree $T_i$ is a shortest path tree in $G_i$.
\end{lemma}

\begin{proof}
  For every vertex $v$ that is not a descendant of $v_{r+1}$ in
  $T_{i-1}$, the path $P$ from $s$ to $v$ in $T_i$ is the same as the
  one in $T_{i-1}$ and hence, a shortest path in $G_i$
  (Lemma~\ref{lem:OldPathInTi}). For all other vertices $v$ in $G_i$,
  it follows from Dijkstra's algorithm that the path from $s$ to $v$
  in $T_i$ is a shortest path.
\end{proof}

Lemmas~\ref{lem:ShortestInGk} and~\ref{lem:TiIsSPTree} together prove
that our algorithm is correct provided it terminates, which we
establish in the next section.

\subsection{Analyzing time and space requirement}
\label{L3:MPDijkC.Analysis}

Although in every iteration we eliminate one exception by modifying
the graph, we introduce copies of certain other exceptions through
vertex replication. Still our algorithm does not iterate indefinitely
because, as we will show in this section, the incremental construction
of the shortest path tree (Sec.~\ref{L3:MPDijk.MakeT}) guarantees that
we do not discover more than one copy of any exception. We first show
that any exception in $G_{i-1}$ has at most two copies in $G_i$
(Lemma~\ref{lem:XCopies}), and then prove that one of these two copies
is never discovered in the future (Lemma~\ref{lem:HiddenX}):

\begin{lemma}\label{lem:XCopies}
  Let $y \neq x_i$ be any exception in $G_{i-1}$. If the last vertex
  of $y$ is \emph{not\/} an intermediate vertex of $x_i$, then $G_i$
  contains exactly one copy of $y$. Otherwise, $G_i$ contains exactly
  two copies of $y$. In the latter case, one copy of $y$ in $G_i$ ends
  at the old vertex $v$ and the other copy ends at the corresponding
  replica $v'$.
\end{lemma}

\begin{proof}
  Let $\pi = (w_1, w_2, \ldots, w_j)$ be a maximal sequence of
  vertices in $y$ that is a subsequence of $(v_{r-l+1}, v_{r-l+2},
  \ldots, v_r)$. Let $w'_j$ be the replica of $w_j$ in $G_i$. We will
  first show that if there is a vertex $v$ in $y$ right after $\pi$,
  then exactly one of the edges $(w_j, v)$ and $(w'_j,v )$ exists in
  $G_i$. Consider the subgraph of $G_i$ induced on the set of replica
  vertices $\{v'_{r-l+1}, v'_{r-l+2}, \ldots, v'_r\}$: this subgraph
  is a directed path from $v'_{r-l+1}$ to $v'_r$, and the only edge
  that goes out of this subgraph is $(v'_r, v_{r+1})$. Therefore, (i)
  when $(w_j,v) = (v_r, v_{r+1})$, $(w'_j,v) \in G_i$ and $(w_j,v)
  \not \in G_i$, and (ii) otherwise, $(w_j,v) \in G_i$ and $(w'_j,v)
  \not\in G_i$.

  Now $G_i$ has exactly two copies of $\pi$: one through the old
  vertices, and another through the replicas. The above claim implies
  that when there is a vertex $v$ in $y$ right after $\pi$, $G_i$ has
  at most one copy of the part of $y$ from $w_1$ to $v$. However, when
  $\pi$ is a suffix of $y$, $G_i$ has both the copies of the part of
  $y$ from $w_1$ to $w_j$. The lemma then follows because any part of
  $y$ that contains no intermediate vertex of $x_i$ has exactly one
  copy in $G_i$.
\end{proof}

\begin{lemma}\label{lem:HiddenX}
  Let $y \neq x_i$ be any exception in $G_{i-1}$ such that the last
  vertex of $y$ is an intermediate vertex $v$ of $x_i$. The copy of
  $y$ that ends at the old vertex $v$ in $G_i$ is not discovered by
  the algorithm in any future iteration.
\end{lemma}

\begin{proof}
  The copy of the path $(s, v_1, v_2, \ldots, v_r)$ through the old
  vertices in $G_i$ contains $v$. Let $P$ be the part of this path
  from $s$ to $v$. Clearly, $P \in T_{i-1}$, and $P$ does not contain
  any exception because the oracle returns the exception with the
  earlier last vertex. So, the way we construct $T_j$ from $T_{j-1}$
  for any iteration $j \ge i$ ensures that $P \in T_j$.

  Let $y_1$ be the copy of $y$ that ends at $v$. Now $y_1$ is not a
  subpath of $P$ because $P$ does not contain any exception. For any
  $j \ge i$, $P \in T_j$, and both $P$ and $y_1$ end at the same
  vertex, therefore $y_1 \not \in T_j$. So, a packet in iteration $j$
  will not follow $y_1$, and $y_1$ will not be discovered in that
  iteration.
\end{proof}

\begin{lemma}\label{lem:Finite}
  The \textbf{while} loop iterates at most $k = |X|$ times.
\end{lemma}

\begin{proof}
  For any iteration $i$, $G_{i-1}$ contains $x_i$, and $G_i$ does not
  contain $x_i$. Every exception other than $x_i$ in $G_{i-1}$ has
  either one or two copies in $G_i$ (Lemma~\ref{lem:XCopies}). By
  Lemma~\ref{lem:HiddenX}, if an exception has two copies in $G_i$,
  only one of them is relevant in the future. Thus the number of
  exceptions effectively decreases by one in each iteration. The lemma
  then follows.
\end{proof}

To determine the running time, observe that the number of vertices
increases in each iteration. However, we run Dijkstra's algorithm on
at most $n$ vertices in any iteration, because the number of replica
vertices added in each iteration is always less than the number of
vertices in the part of the shortest path tree that is carried over
from the previous tree in our incremental use of Dijkstra. Moreover,
we can make sure that Dijkstra's algorithm examines at most $m$ edges
in iteration $i$, by deleting a few more edges from $G_i$ after
performing the graph modification described in
Sec.~\ref{L3:MPDijk.ModifyG}. More precisely, for each old vertex $v
\in \{ v_{r-l+1}, v_{r-l+2}, \ldots, v_r \}$, since the label
(i.e.,~the ``distance'' from $s$) put on $v$ by Dijkstra's algorithm
in the previous iterations remains unchanged later on, we can safely
delete from $G_i$ all the \emph{incoming} edges of $v$ without
affecting future modifications. (Note that for all $j \in [r-l+1, r]$,
old vertices $v_j$ and $v_{j+1}$ are no longer adjacent in $G_i$,
although the edge $(v_j, v_{j+1})$ still exists in $T_i$.) It is not
hard to see that the number of new edges in $G_i$ is now equal to the
number of edges deleted from $G_{i-1}$.

\begin{theorem}\label{thm:Correct}
  The algorithm computes a shortest $X$-avoiding path in $O(k n \log n
  + k m)$ time and $O(n + m + L)$ space.
\end{theorem}

\begin{proof}
  The correctness of the algorithm follows from
  Lemmas~\ref{lem:ShortestInGk} and~\ref{lem:TiIsSPTree}.

  Let $l_i$ be the number of intermediate vertices of the exception
  discovered at the $i$th iteration (thus the size of the exception is
  $l_i + 2$). The $i$th iteration adds $l_i$ vertices. Since the
  algorithm iterates $k$ times (Lemma~\ref{lem:Finite}), there are $n
  + \sum_{i=1}^k l_i < n + L$ vertices in the graph at
  termination. Because in each iteration the number of added edges is
  equal to the number of deleted edges, the space requirement is $O(n
  + m + L)$.
  
  Each iteration of our algorithm takes $O(|V| \log |V| + |E|) = O(n
  \log n + m)$ time, and the total time requirement follows.
\end{proof}

We note that in practice, the algorithm will not discover all $k$ of
the forbidden paths. It will discover only the ones that ``interfere''
in getting from $s$ to $t$.

%

%

\section{Extensions}
\label{L2:MPDijk2}

This section contains: (1) an algorithm to compute shortest paths from
$s$ to every other vertex in $G$; (2) an analysis in the case when $X$
is given explicitly; and (3) a version of the algorithm where the
oracle returns \emph{any\/} exception on a query path, rather than the
exception that ends earliest.

The algorithm in Sec.~\ref{L2:MPDijk} can be extended easily to
compute a shortest path from $s$ to every other vertex in $G$. We
simply repeat the previous algorithm for every vertex in $G$, but with
a small change: in every iteration (except of course the first one) we
use the graph and the shortest path tree constructed at the end of
previous iteration. Since every exception in $X$ is handled at most
once, the \textbf{while} loop still iterates at most $k$ times, and
therefore, the time and space requirements remain the same.

\begin{theorem}\label{thm:Correct2}
  The algorithm computes shortest $X$-avoiding paths from $s$ to all
  other vertices in $O(k n \log n + k m)$ time and $O(n + m + L)$
  space.
\end{theorem}

Our algorithm applies when $X$ is known explicitly; taking into
account the cost of sorting $X$ so that we can efficiently query
whether a path contains an exception we obtain:


\begin{theorem}\label{thm:Correct2.5}
  When $X$ is known a priori, we can preprocess the graph in $O(k n
  \log(k n) + k m)$ time and $O(n + m + L)$ space so that we can find
  a shortest $X$-avoiding path from $s$ to any vertex in $O(n + L)$
  time.
\end{theorem}

Recall that Villeneuve and Desaulniers~\cite{Villeneuve.05} solved
this problem in $O((n + L) \log(n + L) + m + d L)$ preprocessing time,
$O(n + m + d L)$ space and $O(n + L)$ query time. Our algorithm is
more space efficient than theirs. Our preprocessing is slightly slower
in general, although it is slightly faster in the special case $L =
\Theta(k n)$ and $m = o(d n)$ (intuitively, when the
exceptions are long, and the average degree of a vertex is much
smaller than the largest degree).

%
%

Finally, returning to the case where $X$ is not known a priori, we
consider a weaker oracle that returns any exception on the query path,
rather than the exception that ends earliest. At the cost of querying
the oracle more often, we obtain a better run-time. The idea is to
query the oracle \emph{during\/} the construction of a shortest path
tree. The algorithm is very similar to Dijkstra's, the only difference
is that it handles exceptions inside Dijkstra's loop. More precisely,
right after a vertex $v$ is dequeued and added to the current tree, we
try the $s$-$v$ path in the tree. If the path is exception avoiding,
we update the distances of the neighbors of $v$ and go to the next
iteration, as in ``traditional'' Dijkstra's algorithm. Otherwise, we
remove $v$ from the current tree, perform vertex replication and edge
deletion as described in Sec.~\ref{L3:MPDijk.ModifyG}, and then go to
the next iteration.

\begin{theorem}\label{thm:Correct3}
  The algorithm described above computes shortest $X$-avoiding paths
  from $s$ to all other vertices in $O((n + L) \log (n + L) + m + d
  L)$ time and $O(n + m + L)$ space.
\end{theorem}

\begin{proof}
  There are at most $n + L$ vertices in the modified graph in any
  iteration. So, the loop in the modified Dijkstra's algorithm
  executes at most $n + L$ times, and the priority queue holds at most
  $n + L$ entries. Moreover, within Dijkstra's loop vertex replication
  and edge deletion take $O(d L)$ time in total. The running time then
  follows. The proof of correctness is similar to that of
  Theorem~\ref{thm:Correct2} except that the ``current'' shortest path
  tree is no longer a spanning tree in the current graph.
\end{proof}

This new algorithm is faster than the old algorithm of
Theorem~\ref{thm:Correct2} in general but makes as many as $n+L$
queries to the oracle versus at most $k$ oracle queries for the old
algorithm. The old algorithm is slightly faster in the special case $L
= \Theta(k n)$ and $m = o(d n)$.

%

%

\section{Conclusion}
\label{L2:Conclusion}

Motivated by the practical problem of finding shortest paths in
optical networks, we introduced a novel version of the shortest path
problem where we must avoid forbidden paths, but we only discover the
forbidden paths by trying them. We gave an easily implementable,
polynomial time algorithm that uses vertex replication and incremental
Dijkstra.

As we have mentioned before, in practice our algorithms will not
discover all the forbidden paths in $X$. In fact, the running time of
each of our algorithms is determined by only the forbidden paths that
``interfere'' in getting from $s$ to $t$. An interesting open problem
is to bound the number of such paths. We conjecture that in a real
optical network, the number of such paths is $o(k)$, and therefore,
our algorithms run much faster in practice.

\section*{Acknowledgment}
We wish to thank Erik Demaine for useful discussion, and an anonymous
referess for important suggestions.

\bibliographystyle{plain}

\end{document}